\documentclass[twocolumn,showpacs,preprintnumbers,amsmath,amssymb,dvipdfmx]{revtex4-1}

\usepackage[dvipdfmx,hiresbb]{graphicx}
\usepackage{graphicx}
\usepackage{dcolumn}
\usepackage{bm,color}
\usepackage{amssymb}

\newcommand{\be}{\begin{eqnarray}}
\newcommand{\ee}{\end{eqnarray}}
\newcommand{\nn}{\nonumber\\}

\newcommand{\la}{\langle}
\newcommand{\ra}{\rangle}

\def\ss#1{\rm\scriptscriptstyle{#1}}

\begin{document}

\title{Localization and delocalization of fermions in  a background
of correlated spins}

\date{\today}

\author{Tetsuya Takaishi$^1$} 
\author{Kazuhiko Sakakibara$^2$}
\author{Ikuo Ichinose$^3$}
\author{Tetsuo Matsui$^4$}
\affiliation{${}^1$ Department of Physics,
Hiroshima University of Economics, Hiroshima, 731-0192 Japan }
\affiliation{${}^2$ Department of Physics, Nara National College of 
Technology, Yamatokohriyama, 639-1080 Japan }
\affiliation{${}^3$ Department of Applied Physics,
Nagoya Institute of Technology, Nagoya, 466-8555 Japan }
\affiliation{${}^4$ Department of Physics, Kindai University, 
Higashi-Osaka, 577-8502 Japan }


\begin{abstract}
We study the (de)localization phenomena of one-component
lattice fermions in spin backgrounds. 
The O(3) classical spin variables on sites fluctuate thermally
through the ordinary nearest-neighbor coupling.
Their complex two-component (CP$^1$-Schwinger boson) representation  
forms a composite U(1) gauge field on bond, which 
acts on fermions as a fluctuating hopping amplitude in a gauge invariant manner.
For the case of antiferromagnetic (AF) spin coupling, the model has close
relationship with the $t$-$J$ model of strongly-correlated electron systems.
We measure 
the unfolded level spacing distribution of fermion energy eigenvalues
and the participation ratio of energy eigenstates. 
The results for AF  spin couplings suggest a possibility that,
in two dimensions,  all the energy eigenstates are localized.
In three dimensions, we find that there exists a mobility edge, and 
estimate the critical temperature $T_{\ss LD}(\delta)$ of the 
localization-delocalization transition
at the fermion concentration $\delta$. 
\end{abstract}

\maketitle


\section{Introduction}
\setcounter{equation}{0}

Localization-delocalization (LD) transition of fermions
has been one of the most important problems in quantum statistical physics 
and random systems. 
After the seminal work by Anderson~\cite{anderson}, its relation
to the scaling theory, nonlinear sigma models, and the universality class of random matrix theory have been studied extensively~\cite{matrix}. 
The effect of interactions among fermions on the localization has been studied
mostly by using the perturbation theory~\cite{fukuyama}. 
Recently, effects of strong correlations between fermions on a localization 
have attracted interests in modern perspectives.
Among others, Smith et al.\cite{smith1} considered fermions 
in one dimension (1D) coupled with the $z$-component of $s=1/2$ spins, 
and studied many-body localization (MBL) driven by spontaneously generated 
disorders and its relation to disentanglement, etc.
 
As another interesting work, Kovacs et.al.~\cite{kovacs}
studied a  LD-like transition in  quantum chromodynamics (QCD)
by directly analyzing the eigenstates of relativistic kernel of quarks moving 
in gluon backgrounds.
They employed a quenched approximation; i.e.,
gluon configurations are generated by the SU(N) pure gauge theory,
ignoring the back-reaction by quarks.
This system is substantially different from the Anderson model 
on the following points;
(i) the background gluon field acts on fermions not as a random site potential 
but as a fluctuating hopping amplitude on link,
and (ii) the hopping amplitudes are not independent link by link but 
correlate each other via the SU(N) gauge-field dynamics.
QCD is a strong-coupling system due to large fluctuations of gluons,
and it is regarded as a strongly-correlated system, even though there exist no 
self-couplings of quarks.

Recently, there appeared a couple of interesting systems sharing similar properties
to the above mentioned QCD.
The spin model by Kitaev~\cite{kitaev} 
is represented in terms of Majorana fermions moving in a static Z(2) gauge field.
Smith et.al.~\cite{smith2} introduced fermion models coupled with a Z(2) gauge field 
to study the MBL, focusing on the role of local gauge symmetry. 

The purpose of the present work is to study yet another type of random-gauge system;
spinless fermions coupled with a composite U(1) gauge field, which comes from
the Schwinger-boson (CP$^1$) representation of the O(3) spins. 
This gauge theory of fermions is not only simple and universal but also
is closely related with the $t$-$J$ model~\cite{tjmodel} 
of strongly-correlated electron system  in the slave-fermion 
representation~\cite{sf1,sf}. 
Strong interactions between electrons induce fluctuating spin degrees of freedom 
with nontrivial correlations.
In this perspective, the fermions in the present model are nothing but doped holes 
in the spin background.
As far as we know, the problem of localization of doped holes in strongly-correlated
electron systems has not been addressed yet.

In this work, we shall study the system in two and three spatial dimensions
by numerically measuring quantities concerning to the localization phenomena.
These numerical studies show that this system 
exhibits a LD transition in three dimensions (3D) 
for the antiferromagnetic (AF) case, and we estimate the critical temperature 
of the LD transition as a function of the fermion concentration.
Together with the numerical results for the ferromagnetic (FM) case, we 
suggest a coherent structure of the phase diagram of LD transition. 

This paper is organized as follows.
In Sec.~II, we introduce the target model and explain its relation to
the $t$-$J$ model of strongly-correlated electrons.
In the model, a spinless fermion moves in the spin background
with nontrivial O(3) correlations.
We also explain the outline how to prepare the spin configurations
and use them to calculate physical quantities studied in the successive sections. 
In Sec.~III, we show the numerical results of the unfolded level spacing distribution
(ULSD) for the 2D and 3D cases. 
We find that the system-size dependence of ULSD reveals an important difference 
in the 2D and 3D cases, which allows us to
determine the location of the mobility edge for the 3D system.
In Sec.~IV, we investigate the participation ration (PR), which is another 
useful quantity used for the study on LD transitions.
Numerical results verify the observations obtained in Sec.~III.
In Sec.~V, we use these results to locate the LD transition temperature 
as a function of fermion concentration. 
In Sec.~VI, we present the results for the FM spin background,
and discuss a possible behavior of the LD transition temperature.
Section VII is devoted for conclusion.

\section{Model}
\setcounter{equation}{0}

Before going to study on the target model, 
let us consider the Anderson model to fix the notations and concepts
that we will use in the rest of this work.
The Hamiltonian of the Anderson model is given as follows~\cite{anderson};
\begin{equation}
\hat{H}_{\ss A}
=-t\sum_{x,i}\left(\hat{\psi}^\dag_{x+i}\hat{\psi}_x+{\rm H.c.}\right)+\sum_x
V_x\hat{\psi}^\dag_x\hat{\psi}_x,
\label{ha}
\end{equation}
where $\hat{\psi}_x$ is the annihilation operator of fermion 
at site $x$ of the $D$-dimensional lattice without self-interactions,
satisfying $[\hat{\psi}_x,\hat{\psi}^\dag_y]_+=\delta_{xy}.$
$t$ is the hopping amplitude from $x$ to $x \pm i$ with
$i =1,\cdots, D$.  
$V_x$ is a random potential distributing uniformly in $[-\frac{U}{2}, +\frac{U}{2}]$.
We measure the energy eigenvalue $\lambda$ of 
$\hat{H}_{\ss A}$ from the band center ($\lambda=0$).
For $D=2$, all the energy eigenstates for $U > 0$ are
localized, while in 3D,  there exists a mobility edge (ME) at
$\lambda=\pm \lambda_{\ss{ME}}(U)$~\cite{ucrit}.
ME separates 
extended states for  $-\lambda_{\ss{ME}} < \lambda <\lambda_{\ss{ME}}$
and localized states for $|\lambda| > \lambda_{\ss{ME}}$.
Fermi energy $\lambda_{\ss F}(\delta)$ is an increasing function of 
fermion density $\delta\ (\equiv$ average fermion number/site). The critical density $\delta_c$ of the LD transition is
defined by $\lambda_{\ss F}(\delta_c) = -\lambda_{\ss{ME}}$. 
Then as $\delta$ crosses $\delta_c$ from below, there appear extended states. 
If fermions are charged, the system 
changes from an insulator  to a metal at $\delta=\delta_c$.
The energy spectrum of the target model has a similar structure to that 
of the Anderson model.
The primary concern is whether the ME exists or not.

The Hamiltonian of the target model $\hat{H}_\psi$  is given by
\be
\hspace{-0.5cm}
\hat{H}_\psi&=&-t\sum_{x,i}\left(Q_{xi}\hat{\psi}^\dag_{x+i}\hat{\psi}_x
+{\rm H.c.}\right),\nn
 Q_{xi}&\equiv& \bar{z}_x z_{x+i}=\sum_{\sigma=1}^2{z}^*_{x\sigma}z_{x+i,\sigma}.
\label{hpsi}
\ee 
Hereafter, we put $t=\frac{1}{2}$ as the unit of energy. 
The complex fermion hopping amplitude $Q_{xi}$ is defined on each link $(x,x+i)$
in terms of the CP$^1$ variable~\cite{cp1} or Schwinger bosons $z_{x\sigma}$.
$z_{x\sigma}\ (\sigma=1,2)$ is the two-component complex site {\it variables}
on $x$ and satisfies the following local constraint;
\be
z_x&=&(z_{x1},z_{x2})^{\rm t}, \ z_{x\sigma} (\sigma=1,2) \in{\bf C}\nn
\bar{z}_x z_x&\equiv& 
\sum_\sigma z^*_{x\sigma}z_{x\sigma}=1.
\ee
(The bar symbol denotes Hermitian conjugate.)
This $z_x$ forms a background classical O(3) spin vector $\vec{S}_x$ as
\be
\vec{S}_x=\bar{z}_x\vec{\sigma}z_x,\ \vec{S}_x\cdot\vec{S}_x=1,
\label{o3}
\ee
where  $\vec{\sigma}$ are the Pauli matrices. 

The target Hamiltonian $\hat{H}_\psi$ in Eq.~(\ref{hpsi}) can be regarded
as a model describing dynamics of holes doped in many-body spin degrees of freedom
that are generated by the strong correlations between electrons.
A typical model of such strongly-correlated electron systems is the $t$-$J$ 
model~\cite{tjmodel} whose Hamiltonian is given as
\be
\hat{H}_{tJ}&=&-t\sum_{x,i,\sigma}\left(\tilde{C}_{x+i,\sigma}^\dag 
\tilde{C}_{x\sigma}+{\rm H.c.}\right)\nn
&+&J\sum_{x,i}\left(\hat{\vec{S}}_{x+i}\hat{\vec{S}}_x
-\frac{1}{4}\hat{n}_{x+i}\hat{n}_x\right),\label{htj} \\
\tilde{C}_{x\sigma}&\equiv&\left(1- \hat{C}^\dag_{x\bar{\sigma}}
\hat{C}_{x\bar{\sigma}}\right)\hat{C}_{x\sigma}, \label{htj2} \\
\hat{\vec{S}}_x&\equiv& \frac{1}{2}\hat{C}^\dag_x\vec{\sigma}\hat{C}_x,\
\hat{n}_x\equiv \sum_{\sigma} \hat{C}^\dag_{x\sigma}\hat{C}_{x\sigma},
\nonumber
\label{htj1}
\ee
where $\hat{C}_{x\sigma}$ is the annihilation operator of electron at the site $x$  
with the spin $\sigma =1(\uparrow), 2(\downarrow)$, and 
$\bar{\sigma}=1(2)$ for $\sigma=2(1)$. 
The physical states of the $t$-$J$ model exclude the double-occupancy states such as $\hat{C}^\dag_{x\uparrow}\hat{C}^\dag_{x\downarrow}|0\ra$ due to 
the strong electron correlations, i.e, the strong on-site repulsion of electrons.
Due to this local constraint, the specific operators $\tilde{C}_{x\sigma}$ in Eq.~(\ref{htj2})
have been introduce and the hopping term in Eq.~(\ref{htj1}) is expressed 
in terms of them.
The same on-site repulsion also induces the spin-spin interaction in $\hat{H}_{tJ}$ in
Eq.~(\ref{htj}).

The local constraint is faithfully treated by using the slave-fermion 
representation~\cite{sf1,sf}.
In the slave-fermion representation of electron,
$\hat{C}_{x\sigma}$ is represented as a composite,
\be
 \hat{C}_{x\sigma}=\hat{\psi}_x^\dag \hat{a}_{x\sigma},
 \label{csf}
 \ee
where $\psi_x^\dag$ is the creation operator of one-component fermionic hole 
[we use the same notation $\psi_x$ as the fermion operator in 
Eq.~(\ref{hpsi}) because they are to be regarded as the same thing] 
and $a_{x\sigma}$ is the annihilation operator of two-component bosonic 
spin.
The physical states are defined by imposing
the no-double occupancy condition such as,
\be 
\sum_\sigma \hat{a}_{x\sigma}^\dag 
\hat{a}_{x\sigma}+\hat{\psi}_x^\dag \hat{\psi}_x=1.
\label{sfconst}
\ee
We introduce a set of bosonic operators, $z_{x\sigma}$,
and the following parametrization;
\be
\hat{a}_{x\sigma}&=&(1-\hat{\psi}^\dag_x\hat{\psi}_x)^{1/2}\hat{z}_{x\sigma}
\nonumber \\
&=&(1-\hat{\psi}^\dag_x\hat{\psi}_x)\hat{z}_{x\sigma},
\label{asf}
\ee
where we have used the fact that the eigen-values of $\hat{\psi}^\dag_x\hat{\psi}_x$
are $0$ and $1$.
Then,  Eq.~(\ref{sfconst}) is reduced to
the so-called CP$^1$ constraint, 
 \be
 \sum_\sigma
\hat{z}^\dag_{x\sigma}\hat{z}_{x\sigma}=1.
\label{cp1}
\ee 
It is rather straightforward to derive the Hamiltonian $\hat{H}_\psi$ 
in Eq.~(\ref{hpsi}) from the hopping term in $\hat{H}_{tJ}$ in Eq.~(\ref{htj}).
In particular, for the low doping case, where the hole concentration
$\delta =\la \hat{\psi}^\dag_x\hat{\psi}_x\ra$ is small, one may set
$\hat{a}_x \simeq \hat{z}_x$ in Eq.~(\ref{asf}) and the electron hopping term
in Eq.~(\ref{htj}) becomes~\cite{smalldelta} 
\be
\tilde{C}_{x+i,\sigma}^\dag \tilde{C}_{x\sigma}+\mbox{H.c.}
\to
\hat{z}^\dagger_{x+i,\sigma}\hat{\psi}_{x+i}\hat{\psi}^\dagger_x \hat{z}_{x\sigma}
+\mbox{H.c.}
\ee
This is just the hopping term of the Hamiltonian $\hat{H}_\psi$ of  Eq.~(\ref{hpsi}).

In the present work, we treat $\{z_x\}$  in $\hat{H}_\psi$ 
as random variables obeying certain distribution density $\rho(z)$.
Explicitly we take $\rho(z)$ as the Boltzmann distribution 
at the temperature $T$ 
with the energy $E_z$ of the following O(3) spin model;
\be
&&E_z=-J\sum_{x,i} \vec{S}_{x+i}\cdot\vec{S}_x, \nonumber   \\
&&\rho(z)=\frac{\exp(-\beta E_z)}{\int[dz] \exp(-\beta E_z)},
\label{ez}
\ee
where $J$ is the ferromagnetic (FM $J>0$) or AF ($J < 0$) exchange coupling, 
$\beta\equiv 1/(k_{\ss B}T)$, and $[dz]=\prod_xdz_x$ is the Haar measure.
Eq.~(\ref{ez}) is obviously motivated by the spin-spin interactions in the 
$t$-$J$ model in Eq.~(\ref{htj})~\cite{sgnj}.
It should be remarked here that the above treatment of the spin and hole 
variables is motivated by the recent works on MBL for 
systems in which fast and slow particles exist and slow particles serve as a random
quasi-static background potential for fast particles~\cite{Schiulaz,yao}.
Namely, our treatment is viewed as a quenched approximation for
the total Hamiltonian $\hat{H}_{\psi z}=\hat{H}_\psi +E_z$
regarding $\{z_x\}$ as quenched variables and neglecting the back-reaction 
from fermions to spins.
It is quite straightforward to extend the present formalism to other cases of
spin correlation.
In the practical calculation below, we prepare $M (\simeq 1000)$ such $z$-samples, 
each of which  is chosen in every 10 sweeps of the Hybrid Monte Carlo 
simulations.

$\hat{H}_\psi$ of Eq.~(\ref{hpsi}) and $\vec{S}_x$ of Eq.~(\ref{o3})
are invariant under the local U(1) gauge transformation~\cite{cinv}, 
\be
z_{x\sigma}\to e^{i\varphi_x} z_{x\sigma}, \hat{\psi}_x\to e^{i\varphi_x}\hat{\psi}_x,
\ee
where $\varphi_x$ is an arbitrary phase for a local gauge transformation.
This way to couple fermions with O(3)  spins $\vec{S}_x$
by introducing $z_x$ in a gauge-invariant manner reminds us the well known
way to couple charged particles with electromagnetic field by introducing 
gauge field $A_{x\mu}$.
Because $\psi_x$ represents doped holes in Ref.~\cite{sf1, sf}, the present model 
may have some universality common to hole-doped spin models.
In the present work, we study the simplest case in Eq.~(\ref{ez}).

We consider the $D$-dimensional spatial lattice with $N=L^D$ sites and the periodic boundary condition.
For each $z$-sample, we calculate numerically $N$ eigenvalues $\lambda$ 
and eigenfunctions $\phi_x(\lambda)$ of the Hermitian 
fermion kernel $\Gamma$ in $\hat{H}_\psi$,
\be
\hat{H}_\psi&=&\sum_{x,y}\hat{\psi}^\dag_x\Gamma_{xy}(z)\hat{\psi}_y,\nn 
\Gamma_{xy} &=& -\frac{1}{2}\sum_i \bar{z}_y z_x(\delta_{y,x-i}+\delta_{y,x+i}).
\label{gamma}
\ee
In the eigenvalue equation, 
\be
\sum_y\Gamma_{xy} \phi_y(\lambda) =\lambda \phi_x(\lambda),
\ee
$\Gamma_{xy}$ and $\phi_x$ gauge-transform covariantly
as $\Gamma_{xy}\to e^{i\varphi_x}\Gamma_{xy}e^{-i\varphi_y}$ and
$\phi_x\to e^{i\varphi_x}\phi_x$, whereas 
$\lambda$'s are gauge-invariant.
By using $\lambda$ and $|\phi_x(\lambda)|$,  we calculate the ULSD and PR. 
Then, we average these quantities over $M$ $z$-samples.
The obtained results give important informations about (de)localization of fermions
$\hat{\psi}_x$.

To understand the $\beta$-dependence of the results given below, it is useful
to recall that the O(3) spin model (\ref{ez}) has a symmetry under 
AF $\leftrightarrow$ FM with $J\leftrightarrow -J$, and 
the 2D model has only the paramagnetic (PM) phase, whereas
the 3D model exhibits the second-order AF(FM)-PM phase transition at $\beta J \simeq -1.4(1.4)$.
The two amplitudes on the  link $(x,x+i)$, 
(i)  FM spin-pair amplitude $Q_{xi} =\bar{z}_x z_{x+i}$ and 
(ii) AF(resonating-valence bond) amplitude, 
\be
R_{xi}
\equiv z_{x1}z_{x+i,2}-z_{x2}z_{x+i,1},
\ee
satisfy the identity,
\be
|Q_{xi}|^2+|R_{xi}|^2=1,
\ee
and the each term in $E_z$ is expressed as 
$\vec{S}_{x+i}\cdot\vec{S}_x=|Q_{xi}|^2-|R_{xi}|^2=
1-2|R_{xi}|^2=2|Q_{xi}|^2-1$.

As  $\beta J$ increase from $-\infty$ (deep AF phase) to $\infty$ (deep FM phase),
the average squared magnitude $\la |Q_{xi}|^2 \ra$ increases monotonically from 
zero to 1.  
In particular,  in the strong paramagnetic  phase ($|\beta J|\sim 0$),
$\la |Q_{xi}|^2 \ra \simeq \la |R_{xi}|^2 \ra \simeq \frac{1}{2}$.
Generally speaking, as we see below, 
the population of delocalized states (if any) in the whole spectrum
increases as $|Q_{xi}|$ increases. 
\\

\section{Unfolded level spacing distribution (ULSD)}
\setcounter{equation}{0}
\subsection{Definition of ULSD}

The ULSD is one of the commonly used quantities~\cite{matrix, kovacs} 
to find a ME in an 
energy spectrum.  Due to the sublattice symmetry of the system (\ref{gamma}), 
$\lambda$'s distribute  symmetrically around $\lambda=0$, and we focus 
on the half of them ($\lambda \geq 0$). 
For each $z$-sample, we sort
 $N/2$\ $ \lambda$'s as  
$\lambda_m\ (m=1,\cdots, N/2$)  with 
$\lambda_m \leq \lambda_{m+1}$.
We group these $\lambda$'s into $K$ successive sets (cells) 
$C_k (k=1,\cdots,K)$
such that $C_k$ contains $\Lambda_k$ successive $\lambda$'s as 
$C_k=(\lambda^k_1,\cdots,\lambda^k_{\Lambda_k})$,  where
$\lambda^k_\alpha\equiv \lambda_{T_k+\alpha}$ with $T_{k}\equiv\sum_{\ell=1}^{k-1}\Lambda_\ell$.
For each $C_k$, we introduce $\Lambda_k-1$ unfolded level spacings
$s^k_\alpha\ ( \alpha=1,\cdots,\Lambda_k-1)$ defined as
\be 
\hspace{-0.5cm}
s^k_{\alpha}&\equiv& \frac{\lambda^k_{\alpha+1}-
\lambda^k_{\alpha}}{\Delta\lambda^k},\ \Delta\lambda^k \equiv 
\frac{\lambda^k_{\Lambda_k}-\lambda^k_1}{\Lambda_k-1}, 
\ee
where $\Delta\lambda^k$ is the average of the $(\Lambda_k-1)$ nearest-neighbor 
level spacings over $C_k$.

Then we assemble these $s \ (=\{s^k_{\alpha}\})$ for each cell $C_k$ 
over $M$ $z$-samples, and calculate their distribution $P_k(s)$.
By definition, $P_k(s)$ satisfies the following identities;
\be
\int_0^\infty ds P_k(s) =1,\ \ \int_0^\infty ds s P_k(s) = 1
\ee
(owing to the unfolding).
In the following calculations, we choose the parameters as
$K =40, \Lambda_k \simeq N/(2K)$.

In the random matrix theory~\cite{matrix}, typical behaviors of
$P(s)$ ($=P_k(s)$ for each cell $C_k$ in the present context) 
are known as 
\be
&{\rm Poisson\ type}\ & P_{\ss P}(s) \propto  \exp(-s),\nn
&{\rm Wigner\ type}\ &P_{\ss W}(s) \propto  s^c \exp(-d s^2),
\label{ps}
\ee
with certain positive constants $c$ and $d$.
{\em The
behavior of $P_k(s)$ near $s\simeq 0$ clarifies whether the eigenstates 
in the $k$-th cell are localized or extended}, i.e.,
\be
P_k(0) \left\{ \begin{array}{ll}
\neq 0,& {\rm Poisson, \ localized\ state}\\
\simeq 0, & {\rm Wigner, \ extended\ state.} \\
\end{array}
\right.
\label{rule}
\ee
Because  localized eigenstates have a same shape with
a finite extension (localization length) 
and to be distinguished by their locations (centers).
These states are degenerate in energy $(s=0)$, giving rise to $P_k(0)>0$. 
On the other hand, extended eigenstates are made of superpositions of
localized basis states as in Bloch wave so that the degeneracy is 
removed $P_k(0)=0$. 
As $K$ becomes sufficiently large, one can identify the ME
as $\lambda_{\ss{ME}}\simeq \lambda^k_1$ where $P_{k}(0)\sim 0$ and 
$P_{k+1}(0)\neq 0$. 
The {\em whole range of positive $\lambda$ axis} is partitioned into the following sections;
(i) extended states; $0 \leq \lambda \leq \lambda_{\ss{ME}}$,
(ii) localized states; $\lambda_{\ss{ME}} < \lambda\leq \lambda_{\ss MAX}$,
(iii) no states; $\lambda_{\ss MAX} < \lambda$.

\subsection{Results of ULSD}

\begin{figure}[t]
\includegraphics[width=1.01\linewidth]{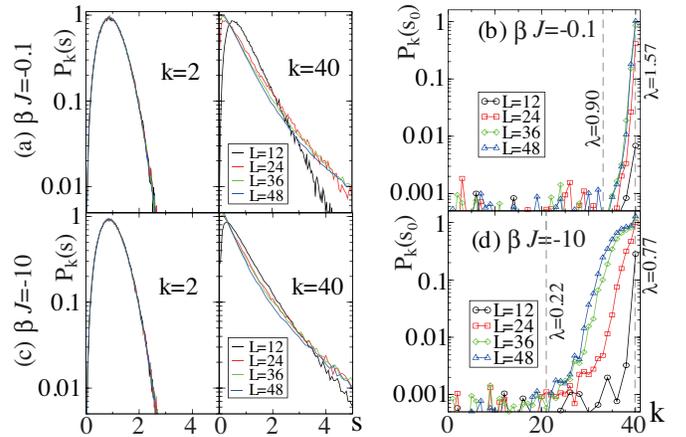}
\vspace{-0.5cm}
\caption{2D AF ULSD $P_k(s)$ ($K=40$) for various $L$'s.
(a) $P_k(s)$ for $k=2$ and 40 ($\beta J = -0.1$).
For $L=12$, both $P_k(s)$'s look the Wigner type. However, 
as $L$ increases, $P_{40}(s)$ approaches to the Poisson type.  
(b) $P_k(s_0)$ at $s_0$, the smallest value of $s$ ($s_0$ is
$0.005\sim 0.015)$.
The region of $k$ for the Poisson type [$P_k(s_0) > 0$] extends to lower  $k$'s
as $L$ increases.
(c), (d) Same plots for $\beta J = -10.0$. 
They show similar but stronger $L$-dependence of $P_k(s_0)$.
The dashed lines in Figs.~(b) (d)  show the locations
of $\lambda_{\ss MAX}(K=40)$ (Right) and
{\it possible} $\lambda_{\ss ME}$ {\it or a crossover 
suggested by the data up to $L=48$} (Left).
}
\label{2Dulsd} 
\end{figure}

Let us turn to numerical results.
In Fig.~\ref{2Dulsd}, we show 2D AF ULSD, $P_k(s)$,
for various linear lattice sizes $L$.
Fig.~\ref{2Dulsd}(a) for $\beta J =-0.1$ shows that 
$P_k(s)$'s with both $k=2$ and $k=40$ look the Wigner type for $L=12$.
However, the curve for $k=40$
seems to approach to the Poisson type as $L$ increases.
In Fig.~\ref{2Dulsd}(b), we show $P_k(s_0)$, where $s_0$ is   
the smallest value of $s$ in each cell $C_k$.
$P_k(s_0)$ shows that this $L$-dependence is systematic. 
That is, as $L$ increases, the region of $k$ with the Poisson type increases 
toward the lower $k$ monotonically.
Figs.~\ref{2Dulsd}(c) and (d) for $\beta J =-10.0$ slightly amplify this $L$-dependence
as expected because localization is more favored than $\beta J =-0.1$. 
This peeling-off phenomenon continues down to some value $k=k_c$  in $L\to \infty$, 
and then $k_c$ is the ME, although the precise determination of $k_c$ 
by $P_k(s_0)$ alone requires scaling arguments using the date for larger $L$'s. 
We shall discuss the ME by using more efficient methods below.

\begin{figure}[t]
\includegraphics[width=1.0\linewidth]{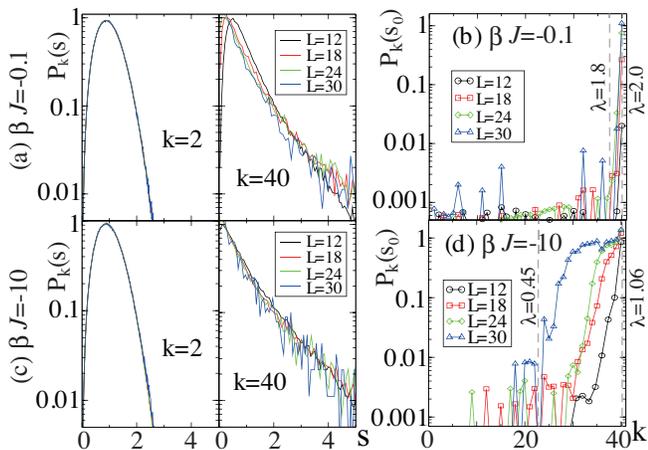}
\vspace{-0.5cm}
\caption{
\!3D AF ULSD $P_k(s)$ ($K\!=\!40$) for various $L$'s.
\!\!\!(a) $P_k(s)$ for $k=2$ and 40 ($\beta J = -0.1$).
As $L$ increases, $P_{40}(s)$ approaches to the Poisson type, while 
$P_2(s)$ remains in the Wigner type. (b) $P_k(s_0)$ ($\beta J = -0.1$).
Its  high-value region looks narrower
than the 2D case of Fig.~\ref{2Dulsd}(b).
(c),(d) Same plots for $\beta J = -10.0$. $P_k(s_0)$ there separates
its high-value region and low-value region at $k\sim 23$
more sharply than Fig.~\ref{3Dulsd}(d) of the 2D case with long tails, 
suggesting a ME. For the dashed lines, see Fig.~\ref{2Dulsd}. }
\label{3Dulsd} 
\end{figure}

In Fig.~\ref{3Dulsd}, we show the results of 3D AF ULSD for $\beta J =-0.1$, and $-10.0$.
Fig.~\ref{3Dulsd}(a) shows that, as $L$ increases,  
$P_{2}(s)$ remains the Wigner type, whereas 
$P_{40}(s)$ changes to the Poisson type. 
$P_k(s_0)$ in Fig.~\ref{3Dulsd}(d) for $\beta J = -10.0$
shows that the regions of lower and higher value of $P_k(s_0)$ are 
more clearly separate than the 2D case of Fig.~\ref{2Dulsd}(d). 

The above observation by using the ULSD seems to indicate that,
both in the 2D and 3D cases, {\it a LD transition or a crossover}  
takes place at finite $\lambda_{\ss{ME}}$.
As mentioned, we shall see that further analyses below
provide us with certain signals for differences in the LD properties of the 2D and 3D systems.

\subsection{ULSD and Mobility edge}

A way to determine the location of ME, $\lambda_{\ss{ME}}$, in a systematic 
manner was proposed in Refs.~\cite{kovacs,integral}. 
It uses the following integral;
\be
I_k=\int_0^{\bar{s}} ds P_k(s), 
\label{ik}
\ee
where we put $\bar{s}\simeq 0.508$~\cite{508} in the practical calculation
following Refs.~\cite{kovacs,integral,196}, but the qualitative results are the same 
for other values of $\bar{s} (\approx 0.5)$.  
Physical meaning of the integral $I_k$ is the following.
For small $\bar{s}$, $I_k$ picks up the behavior of $P_k(s)$ 
in the regime $s\sim 0$, so $I_k$ is small for the Wigner distribution and
large for the Poisson distribution.   
Let us regard $I_k$ as a function
of the smallest $\lambda$ in $C_k$, i.e., $\lambda^k_1$ and define
$I(\lambda^k_1) \equiv I_k$.
For large $K$, $\lambda^k_1$ becomes sufficiently dense, and so
$I(\lambda^k_1)$ becomes a smooth function $I(\lambda)$,
which equals $I_k$ for $\lambda=\lambda^k_1$.
As we mentioned above, one can determine $\lambda_{\ss ME}$,
from the behavior of $I(\lambda)$. Explicitly, we compare below
$I(\lambda)$ with its three typical values;
\be
I_{\ss W} &\simeq 0.12&\ {\rm for}\ P_{\ss W}(s),\nn
I_{\ss P} &\simeq 0.40&\ {\rm for}\ P_{\ss P}(s),\nn
I_c &\simeq 0.196&\ {\rm for}\ P_c(s),
\label{3i}
\ee
where the first two correspond to the Wigner and Poisson distributions of 
Eq.~(\ref{ps}), respectively~\cite{integral,508}.
On the other hand, $P_c(s)$ is the critical distribution 
at the transition point between them~\cite{integral,196}.

\begin{figure}[t]
\includegraphics[width=1\linewidth]{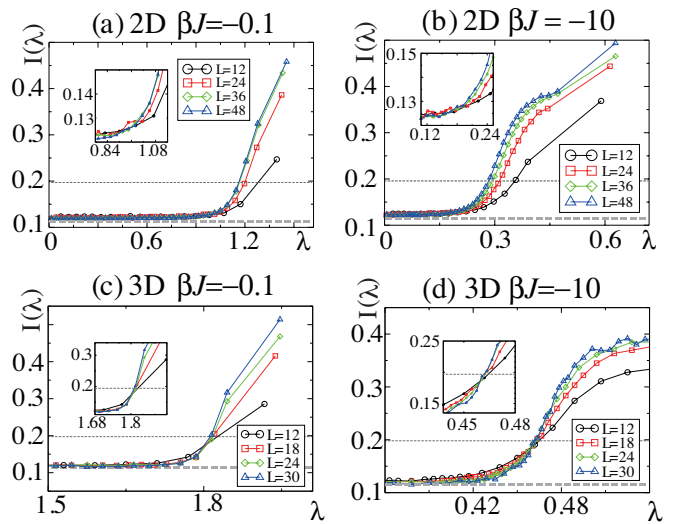}
\caption{$I(\lambda)$ of Eq.(\ref{ik})  for $\beta J = -0.1, -10.0$. 
(a), (b) 2D $I(\lambda)$.
For both $\beta$'s, as $L$ increases, $I(\lambda)$ peels off upward monotonically.
(c), (d) 3D $I(\lambda)$. For both $\beta$'s,  four curves
with different $L$'s cross approximately at a single point in contrast with the 2D case.
This almost $L$-independent point is a candidate of ME.
The lower dashed lines show $I_{\ss W}$ and the upper dashed lines 
 show $I_c$ of the critical statistics $P_c(s)$.}
\label{figik}
\end{figure}

In Fig.~\ref{figik}, we show $I(\lambda)$ for $\beta J= -0.1$, and $-10.0$
and various $L$'s.
In Figs.~\ref{figik}(a) and (b), 2D $I(\lambda)$ remains 
near the value $I_{\ss W}$ in the  lower-$\lambda$ region, 
and it starts to deviate upward as $\lambda$ increases.
As $L$ increases,  the point of deviation shifts to lower $\lambda$'s monotonically.

In Figs.~\ref{figik}(c) and (d), 3D
$I(\lambda)$ increases toward $I_{\ss P}$ as $\lambda$ increases, 
as expected~\cite{ipoisson}. 
The four curves of $I(\lambda)$ for four $L$'s show systematic size-dependence
such that the transition gets sharper as $L$ increases, and interestingly enough,
{\em these four curves seem to cross with each other almost at the same point}
[Insets of Figs.~\ref{figik}(c) and (d)].  
{\em This is in sharp contrast with the 2D case.}
This crossing point is $L$-independent and to be a candidate for the ME
in the infinite-volume limit.
Another candidate of the ME is given by  
$I(\lambda_{\ss ME}) = I_{\rm c}$ for the critical statistics
at the transition point  between the Poisson and the Wigner distributions,
although the critical statistics itself of the present model may 
differ from $P_c(s)$ due to the correlated hopping.
It is interesting that
the above two methods give almost the same estimation of $\lambda_{\ss{ME}}$.

\section{Participation ratio (PR)}
\setcounter{equation}{0}

Another quantity that we use to study the LD transition
is the PR~\cite{kovacs,prn}, which is defined by using a normalized 
eigenfunction $\phi_{x}$ of eigenvalue $\lambda$ as follows; 
\be
{\rm PR}(\lambda) = \frac{1}{\sum_x|\phi_{x}|^4\cdot N}.
\label{defpr}
\ee
To see typical behavior of the PR,
let us calculate PR of a state with $|\phi_{x}|^2=$ constant
on $S$ sites and $\phi_{x} =0 $ on the other $(N-S)$ sites;
\be
\hspace{-0.5cm}\sum_x|\phi_{x}|^2&=&S\cdot|\phi_{x}|^2 = 1,\
{\rm PR} =\frac{1}{S \cdot S^{-2}\cdot N}=\frac{S}{N}.
\label{const}
\ee
Eq.~(\ref{const}) shows that ${\rm PR}$ is the ratio of numbers of participated
(occupied) sites and the total sites.

\begin{figure}[t]
\includegraphics[width=1\linewidth]{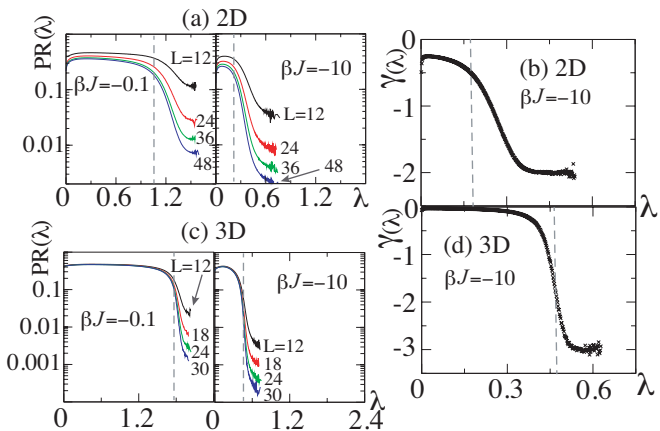}
\vspace{-0.3cm}
\caption{2D and 3D AF PR($\lambda$). 
(a) 2D PR($\lambda$) for $\beta J=-0.1$ and $-10.0$.
(We plot it only for $0 \leq \lambda$ because PR$(\lambda)$ is an even function).
In the small-$\lambda$ region,
PR is fairly large corresponding to extended states, and as $\lambda$ increases,
PR decreases rapidly toward localized states. 
As $L$ increases, PR's for both $\beta J$ reduce monotonically
in the entire region. 
(b) 2D exponent $\gamma(\lambda)$ in the fitting PR($\lambda$) = 
$C L^\gamma$. 
$\gamma (\lesssim -0.3)$ is negative for all $\lambda$, suggesting that PR $\to 0$ as $L \to \infty$.
For large $\lambda$'s $\gamma$ converges to $-2(=-D)$
corresponding to localized states (See the text).
The dashed lines in (a),(b) locate approximate {\it crossover} points.
(c) 3D AF PR($\lambda$) for $\beta J=-0.1$ and $-10.0$. 
In contrast with the 2D PR of Fig.(a),  $L$-dependence 
of PR in the small-$\lambda$ region is quite weak.
(d) 3D exponent $\gamma(\lambda)$. It stays around zero for small $\lambda$'s
and converges to $-D=-3$ for large $\lambda$'s.  
Figs.~(c) and (d) suggest ME around the dashed lines.
}
\label{pr} 
\end{figure}

In Fig.~\ref{pr}(a), we show the averaged value of
2D AF PR($\lambda$) over $M$ $z$-samples for $\beta J= -0.1$ and
$-10.0$.
As $\beta$ increases, the width of PR [the range of $\lambda =
(-\lambda_{\ss MAX},\lambda_{\ss MAX})]$ decreases reflecting the fact that
the magnitude of the FM hopping amplitude $Q_{xi}$ reduces as the AF correlation
increases.
Although  PR exhibits a crossover between high and low-value regions
suggesting existence of a finite $\lambda_{\ss{ME}}$, PR decreases monotonically
as $L$ increases {\em for all $\lambda$'s}.
To study the $L$-dependence of PR($\lambda)$ systematically, 
we fit the obtained data of PR as 
\be
{\rm PR}(\lambda)= C L^{\gamma(\lambda)}.
\ee
In Fig.~\ref{pr}(b), we show the exponent $\gamma(\lambda)$. 
It shows that $\gamma$ is negative $(\lesssim -0.3)$ for all $\lambda$'s, 
suggesting that the $L$-dependence is strong enough,
and all the states get localized (PR$\to 0$) as $L\to \infty$.
If the state consists of a single localized region with finite localization length $\ell$, 
$S$ in Eq.~(\ref{const}) is $\propto \ell^D\sim L^0$ and $\gamma\to -D$ as 
$L\to \infty$~\cite{b}.
$\gamma$ in Fig.~\ref{pr}(b) certainly converges to this value
for $\lambda \gtrsim 0.4$.

In Fig.~\ref{pr}(c), we show 3D AF PR($\lambda$) for $\beta J=-0.1$ and 
$-10.0$. 
In contrast with the 2D PR in Fig.~\ref{pr}(a),  the $L$-dependence 
in the small-$\lambda$ region is very weak and the location of the sharp reduction
at $\lambda \simeq 0.47$ is stable, suggesting existence of a finite ME in the 3D case.
In Fig.~\ref{pr}(d) we show the exponent $\gamma$.
For $\lambda \lesssim 0.3$, $\gamma$ remains almost vanishing,
suggesting the states in that regime are delocalized, and 
around $\lambda\simeq 0.47$, $\gamma$ shifts quickly to $\simeq -D$ of localized states. 
Therefore we conclude that there is a ME for $\beta J = -10.0$ 
at $\lambda_{\ss ME} \simeq 0.47$, and similarly  $\lambda_{\ss ME}
\simeq 1.8$ for $\beta J = -0.1$ [See the dashed lines in Figs.~\ref{pr}
(c) and (d)].
These values are in good agreement with the estimation by using $I(\lambda)$ of 
Figs.~\ref{figik}(c) and (d).
We note that the 3D Anderson model has a similar behavior of PR$(\lambda)$, i.e., 
it has strong depression for localized states as $L$ increases, whereas 
it has almost no $L$-dependence for extended states.

Let us summarize the $L$-dependence of the AF 2D and 3D systems.
Although $P_k$ in Figs.~\ref{2Dulsd} and \ref{3Dulsd} show weaker signals, all the
quantities, $P_k(s), I(\lambda)$, PR have similar behavior, i.e., 
the 2D system shows monotonic $L$-dependence, while the 3D system has 
some fixed point in $\lambda$ that indicates the existence of a ME. 
If the 2D monotonic behavior
continues down to $\lambda\to 0$ as $L\to \infty$, all the states are to be 
localized.

\begin{figure}[b]
\vspace{-0.3cm}
\includegraphics[width=1.00\linewidth]{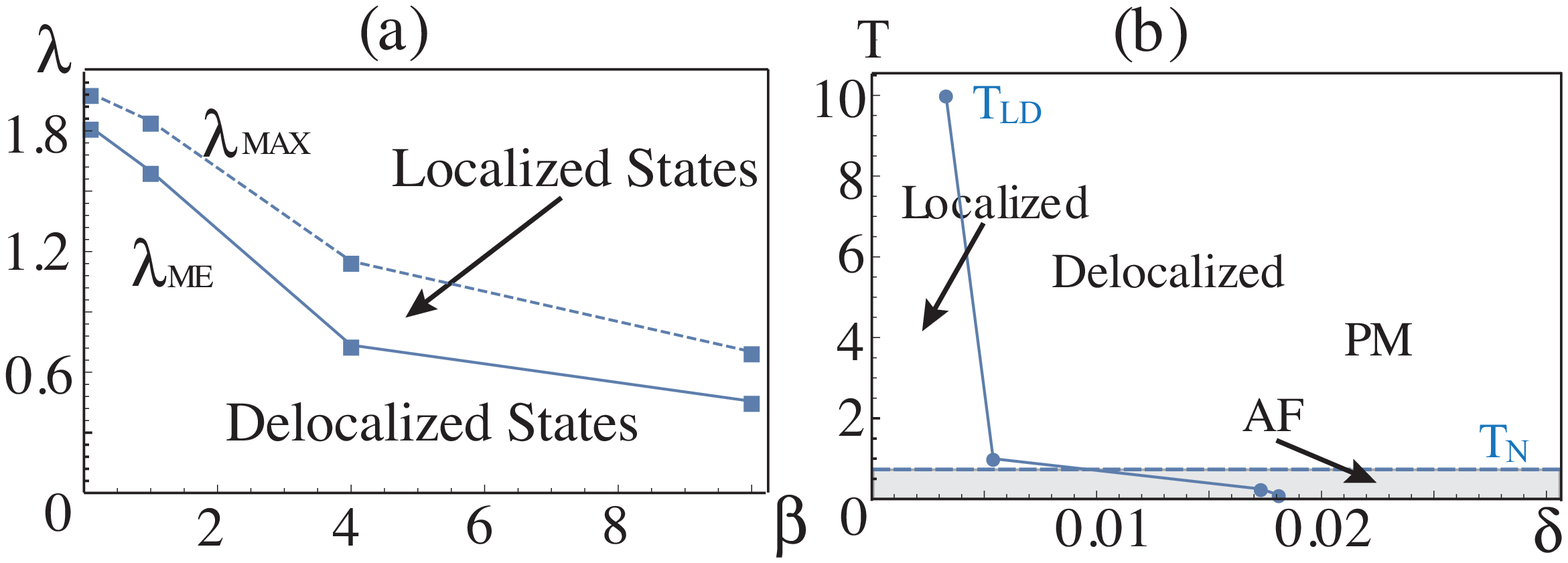}
\vspace{-0.5cm}
\caption{ 
(a) 3D ME. $\lambda_{\ss ME}$ and  
$\lambda_{\ss MAX}$  
vs. $\beta$ (we set $J=-1$).
$\lambda_{\ss ME}$ are determined by $I(\lambda)$ of Fig.~\ref{ik}(c,d)
and similar data for other $\beta$'s.
(b)  3D $T_{\ss LD}(\delta)$ vs. $\delta-T$.
The dashed line shows the N\'{e}el temperature $T_{\ss N}
\simeq 1/1.4$ of the O(3) model  of Eq.~(\ref{ez}). 
In the AF phase, the region of localized states increases.
}
\label{dc} 
\end{figure}


\section{Critical temperature of the LD transition}
\setcounter{equation}{0}

From the numerical studies explained in the previous sections,
we can estimate the LD transition temperature as a function of the fermion
concentration $\delta$.
This problem is addressed in this section.

As the fermion concentration $\delta$ increases from zero
to the critical density $\delta_{\rm c}$,
Fermi energy increases from $-\lambda_{\ss MAX}$ 
to $-\lambda_{\ss ME}$. 
Because $\lambda_{\ss ME}$ depends on $T$ in the present AF 3D system
through Eq.~(\ref{ez}),  
$\delta_{\rm c}$ is determined as a function of $T$, i.e., we have
the critical temperature of the LD transition, $T_{\ss LD}(\delta)$.
To estimate $T_{\ss LD}(\delta)$,  we first show 
$\lambda_{\ss{ME}}$ as a function of $\beta J$ in Fig.~\ref{dc} (a), which is
determined by $I(\lambda)$ of Fig.~\ref{figik} and the PR of Fig.~\ref{pr} for
various $\beta J$. 
To relate $\lambda_{\ss{ME}}$ and $\delta_c$ at sufficiently low $T$'s,
we obtain $\lambda(\delta)$ just by counting the number of 
states with the eigenvalue between $(-\lambda_{\ss{MAX}}, -\lambda)$.
In Fig.~\ref{dc}(b), we plot $T_{\ss LD}(\delta)$ in the $(\delta-T)$-plane.
The slope of $T_{\ss LD}(\delta)$ reduces drastically 
as it crosses the N\'{e}el temperature from above. It may reflect the fact that
$Q_{xi}$ fluctuates more in the AF phase than in the PM  and FM phases, thus 
favoring localized states. 
Fig.~\ref{dc}(b) should be compared with the phase diagrams  
of  various strongly-correlated systems including high-$T$ 
superconductors~\cite{sf, hight, im2}.
In particular, the observation of the enhancement of the localization in the N\'{e}el state 
compared with the paramagnetic state verifies the validity of the present study.


\section{FM coupling}
\setcounter{equation}{0}

\begin{figure}[b]
\begin{center}
\includegraphics[width=1\linewidth]{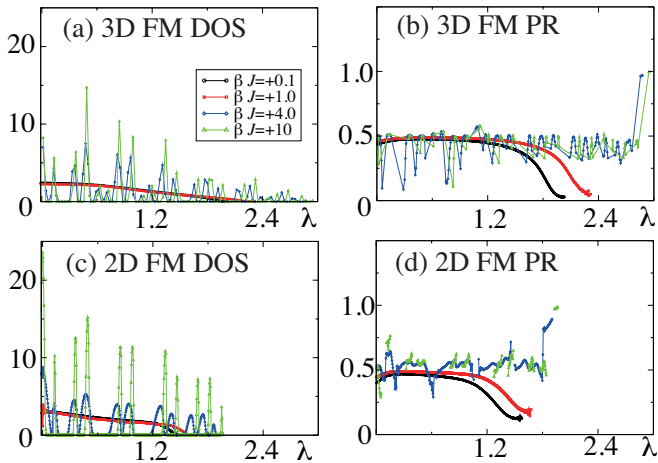}
\end{center}
\caption{(a) DOS $D(\lambda)$ and (b) PR$(\lambda)$ of 
the 3D FM system for $L=12$ and $0.1 \leq \beta J \leq 10$.
$D(\lambda)$ is smooth for  $\beta J = 0.1, 1.0$, whereas it
is spiky for $\beta J = 4.0, 10.0$. 
For $\beta J = 0.1, 1.0$, PR$(\lambda)$ decreases as $\lambda$ approaches to
the band edges. 
In contrast, for $\beta J = 4.0, 10.0$,
PR$(\lambda)$ increases sharply near the band edges.
(c) DOS $D(\lambda)$ and (d) PR$(\lambda)$ of 
the 2D FM system for $L=12$ and $0.1 \leq \beta J \leq 10$.
These quantities exhibit similar behavior with those in the 3D case.
}
\label{prfm}
\end{figure}

The target system in the FM case ($\beta J > 0$) may exhibit qualitatively different 
behaviors from the AF case because $\la |Q_{xi}|^2 \ra$ increases from 0 to 1 as
$\beta J$ varies from $-\infty$ to $\infty$.
In Figs.~\ref{prfm} (a),(c), we show FM density of state (DOS), 
$D(\lambda)$ defined by
\be
D(\lambda)=\frac{d\delta}{d\lambda},
\ee
for $L=12$ and various $\beta J$'s.
In Figs.~\ref{prfm} (b),(d), we show the corresponding PR's. 

Let us see the 3D case first.
In the $\beta J = 0.1$ and $1.0$ cases, which correspond to the PM phase
($0 < \beta J \lesssim  1.4$), 
both $D(\lambda)$ and PR$(\lambda)$ are rather smooth, and PR 
of Fig.~\ref{prfm} (b) has a similar behavior to PR of Fig.~\ref{pr} (c)
for $\beta J = -0.1$ in the PM phase. 
However as $\beta J$ increases into the FM phase ($\beta J \gtrsim 1.4$),  
the FM DOS and PR's for $\beta J =4.0$ and $10.0$ in Figs.~\ref{prfm} (a),(b)  
develop spiking structures, while
the AF PR for $\beta J = -10.0$ in Fig.~\ref{pr} (c) remains smooth.
This reminds us of the plane wave state of a free fermion
with a uniform hopping ($Q_{xi} \to q = |q|e^{i\alpha}$).
The energy $\lambda_{\ss free}$ and $\phi^{\ss free}_x$ of a free fermion  are
given by 
\be
\lambda_{\ss free}&=&- |q| \sum_{i=1}^D\cos(k_i+\alpha),\ k_i=\frac{2\pi n_i}{L},
n_i=1,\cdots,L,\nn
\phi^{\ss free}_x& =&\frac{1}{\sqrt{L^D}}\exp(i\sum_i k_ix_i).
\label{free}
\ee
$\lambda_{\ss free}$ exhibits similar spikes in $D(\lambda)$  
in its interval $\lambda_{\ss free} =(-D|q|,D|q|)$ and PR$(\lambda)$ = 1.
Its origin is just the discreteness of the momentum $k_i$.

For sufficiently large $\beta J$, this similarity can be understood as follows;
The O(3) model of Eq.~(\ref{ez}) for $\beta J \gg 1$ forces $|Q_{xi}|\simeq 1$.
This implies $Q_{xi}\simeq \exp(i\varphi_x)\exp(-i\varphi_{x+i}) $
for each $z$-sample, and furthermore the phase $\varphi_x$ can be absorbed into 
$\hat{\psi}_x$ giving rise to the free fermion system with $q=1$.
However,  Fig.~\ref{prfm} (b) shows  PR $\lesssim 0.5$ instead of PR $\simeq 1$
except for the states in the vicinity of the band edge.
This unexpected reduction of PR by  factor $\sim 2$ 
is attributed to the {\it slight} deviation $1-\la |Q_{xi}|^2\ra \neq 0$
for $\beta J < \infty$ [we estimated it as $1-\la |Q_{xi}|^2\ra \simeq 0.03$ 
for $\beta J = 10.0$].
Although it generates certainly a small change in thermodynamical 
quantities such as the internal energy, 
it induces interference effect on the wave function and the 
energy of fermions in destructive manner.
Similar discrepancy is observed in the random-phase hopping 
model (RPHM)~\cite{rphm}, although the RPHM does not have the local gauge
symmetry.

Let us turn to the 2D case.
In Fig.~\ref{prfm}, we show (c)  2D FM $D(\lambda)$ and (d) 2D FM PR$(\lambda)$
for $L=12$ and various $\beta J$'s.
Both the DOS and PR  
are smooth for smaller $\beta J$ (= 0.1, 1.0) and 
exhibit a spike structure for larger $\beta J$ (= 4.0, 10.0) as in the 3D case.
Explicitly, for all $\beta J$'s, PR$(\lambda)$ keeps $\sim 0.5$ in the central region.
For  $|\lambda|\gtrsim 0.2$, PR$(\lambda)$  decreases rather sharply
for $\beta J = 0.1, 1.0$,
whereas it keeps the similar values and even approaches $\sim 1$
in the vicinity of the band edges for $\beta J = 4.0$ and $10.0$.
This behavior is quite similar to the 3D $D(\lambda)$ of Fig.~\ref{prfm} (b).
We expect that, in the limit $\beta  J \to \infty$, the 2D eigenstates
converge to the plane-wave states of Eq.~(\ref{free}).   

These contrasting behaviors (i.e., smooth and spiky) of the 2D and 3D 
PR$(\lambda)$ near the band edges 
for two regions of $\beta J$, (1) $\beta J \in  (-10.0, +1.0)$ 
and (2) $\beta J \in (4.0, 10.0)$  imply that some critical
value $\beta J = (\beta J)_{\rm c}$ exists at which all the states become delocalized,
i.e., $\lambda_{\ss ME}=\lambda_{\ss MAX}$.
This may be expected as one extends Fig.~\ref{dc} (a) for $J =-1$ 
into the positive-$J$ region.
This critical point is to be induced because
the squared hopping amplitude 
$|Q_{xi}|^2$ runs from 0 to 1 as $\beta J $ runs from $-\infty$ to $\infty$ 
both for the 2D and 3D systems. 
The 2D O(3) spin model has no phase  transitions
in contrast with the 3D model. 
Therefore the phase transition of the correlated-spin background
itself is {\it not} a necessary condition for the existence of 
$(\beta J)_{\rm c}$ itself. 
Calculation of $(\beta J)_{\rm c} \in (1.0, 4.0)$ requires further 
analyses of PR$(\lambda)$, etc.


\section{Conclusion}
\setcounter{equation}{0}

In summary, we studied the LD of a realistic gauge model of  
fermions in the correlated spin background by using the conventional techniques 
for random systems and level statistics.
As emphasized in introduction, the strong correlations between the original
electrons generate the fluctuating spin background with the correlation of
the O(3) model in the present model.
We assume that the spin serves as a random quasi-static background
controlling the fermion hopping amplitude.

First, we studied the model in 2D and 3D for the AF spin coupling by the ULSD.
Finite-size scaling analysis of the ULSD indicates the existence of the ME in the 3D case,
whereas it does not give a clear conclusion for the 2D case.
Then, we investigated the PR and its finite-size scaling.
The results imply that all the states are localized in the 2D case, but more detailed
study is required to obtain a clear conclusion.
In fact, for some related models of a 2D electron gas in a random magnetic field
{\em and} an on-site random potential, a Kosterlitz-Thouless-type
metal-insulator transition was pointed out~\cite{xie} and also existence of
a hidden degree of freedom was suggested~\cite{nguyen}.
{\em As the amplitude and phase of the hopping $Q_{xi}$ are both
random variables}, the model in the present work may have some resemblance with
the above ones.
This is a future problem.
In any case, these methods work well allowing us to
calculate the 3D critical temperature $T_{\rm LD}(\delta)$
of the LD transition. It shows that the region of localized states is enhanced in the 
AF phase. The result of PR for the FM spin coupling 
indicated some critical point $(\beta J)_{\rm c}$ at which all the states
become delocalized ($\lambda_{\ss ME}=\lambda_{\ss MAX}$).
  
Concerning to the relation between the magnetic phase transition
of the  O(3) spin model and the LD phase transition, one might
expect some strong correlation between them.  
In fact, our result that a ME exists in the 3D AF case while
no clear evidence of ME (probably a crossover) in the 2D AF case
is compatible with the fact that the O(3) transition existing in the 3D case
disappears in the 2D case. 
However, we think that this is just accidental coincidence. 
In fact, Anderson model and related models with
uncorrelated randomness 
show a ME in 3D but not in 2D, which is explained without additional 
phase transitions. 
Also our result of Fig.~\ref{dc}(b) shows that
the ME  generally takes place not on the O(3) transition line.
The O(3) spin transition is a thermodynamic transition concerning to a global
change in nature of the system, 
while the LD transition is related with the transport properties,
and the details of each eigenstate are an essential ingredient for that.
This point shares some common aspect with the discussion 
at the end of Sec.~VI for 
the critical value at which $\lambda_{\ss ME}=\lambda_{\ss MAX}$.  
\\

\vspace{1cm} 

\section*{Acknowledgments}
We thank Drs. Shinsuke Nishigaki and Hideaki Obuse for discussion on the random matrix theory and related topics.
Numerical calculations for this work were carried out at the
Yukawa Institute Computer Facility and at the facilities of the Institute of Statistical Mathematics.



\begin{references}
\bibitem{smalldelta}
The case of moderate doping is discussed comprehensively  in Ref.~\cite{sf}.
\bibitem{sgnj}
In $E_z$ of Eq.~(\ref{ez}), we flipped the signature of $J$
from the AF spin coupling of the $t$-$J$ model  (\ref{htj2}).
\bibitem{cinv}
The electron operator 
$\hat{C}_{x\sigma}$ in Eq.~(\ref{htj2}) is also invariant under the
corresponding local gauge transformation, $\hat{a}_{x\sigma}\to 
e^{i\varphi_x}\hat{a}_{x\sigma}, \hat{\psi}_x\to 
e^{i\varphi_x}\hat{\psi}_x.$
\bibitem{508}
The value $\bar{s}=0.508$ is chosen so that the difference between 
two intermediate integrals of 
ULSD, $\int_0^{\bar{s}} ds P(s)$,  for the Poisson type   and the
Wigner type ($c=2, d=4/\pi$) in Eq.~(\ref{ps})
becomes maximum.
\bibitem{ipoisson}
Both in the 2D and 3D cases, $I(\lambda)$ exceeds $\int_0^{\bar{s}}\exp(-s)\simeq 0.40$
for some $\lambda$'s, which implies that $P_k(s)$ there favors localized states 
more than the Poisson type.
\bibitem{b}
If such regions with each volume $\ell^D$ are assembled quasiperiodically
and superposed to make a Bloch-type delocalized state, $S\propto\ell^D (L/\ell)^D$ and $\gamma \propto D/L \to 0 $.
\end{references}
\end{document}